\documentclass[9pt, sigconf, nonacm]{acmart}

\AtBeginDocument{%
  }
 
\setcopyright{none}
\acmDOI{}
\acmISBN{}
\acmBooktitle{}
\acmConference[ICCAD '26]{International Conference on Computer-Aided Design}{November 8-12, 2026}{San Jose, CA, USA}

\usepackage{pgfplots}
\usepackage{algorithm}
\usepackage{algorithmic}
\usepackage{graphicx}
\usepackage{textcomp}
\usepackage{xcolor}

\usepackage{booktabs} 
\usepackage[caption=false]{subfig}
\usepackage{fancyvrb} 
\usepackage{multirow}
\usepackage{diagbox}
\usepackage{tikz}
\usepackage{numprint}
\usepackage{url}
\usepackage{colortbl}
\usepackage{dhucs} 
\usepackage[mathcal]{euscript}

\usepackage{enumitem} 
\usepackage{makecell} 

\usepackage[font=small]{caption}
\usepackage{comment}
\usepackage{lineno}
\usepackage{color} 
\usepackage{setspace} 
\usepackage{relsize}
\usepackage{tabularx}
\usepackage{bm}
\usepackage{hhline}
\usepackage{xurl}

\usepackage{ragged2e}
\usepackage{stfloats}

\setcitestyle{numbers,sort&compress}

\newcommand\refFigure[1]{Figure~\ref{#1}}
\newcommand\refTable[1]{Table~\ref{#1}}

\newcommand\korean[1]{}
\newcommand\blind[1]{XXXX}

\newcommand\LWJ[1]{\textcolor{black}{#1}}

\newcommand\LKW[1]{\textcolor{black}{#1}}
\newcommand\PSC[1]{\textcolor{black}{#1}}
\newcommand\MDJ[1]{\textcolor{black}{#1}}

\renewcommand{\baselinestretch}{1}

\begin{document}
\title{HSF-S: Speed-Optimized Compilation and Acceleration for Hybrid Schr\"odinger--Feynman Quantum Circuit Emulation}

\AtBeginDocument{
  \setlength\abovedisplayskip{2pt}       
  \setlength\belowdisplayskip{2pt}       
  \setlength\abovedisplayshortskip{0pt}  
  \setlength\belowdisplayshortskip{0pt}  
}

\author{Sechan Park, Kyeongwon Lee, Mundo Jeong, Chaebin Jung, and Woojoo Lee}
\affiliation{%
  \institution{\vspace{0.3em} Department of Intelligent Semiconductor Engineering, Chung-Ang University, Seoul, Korea \\ \vspace{0.1em}
}
  \city{}
  \country{} 
}
\email{}
\thanks{This paper has been accepted for publication in the Proceedings of the ACM/IEEE International Conference on Computer-Aided Design (ICCAD), 2026.}

\thanks{
This work was supported in part by the National Research Foundation of Korea (NRF) grant funded by MSIT (No. RS-2024-00345668), and in part by Institute of Information \& communications Technology Planning \& Evaluation (IITP) grants funded by MSIT (No. RS-2023-00277060).

Sechan Park and Kyeongwon Lee contributed equally to this work. 

Woojoo Lee is the corresponding author.}

\begin{abstract}

Hybrid Schr\"odinger--Feynman (HSF) simulation offers an attractive memory--path tradeoff for exact quantum-circuit emulation, but its practical runtime is often dominated by exponential path growth from cross-boundary two-qubit gates. 
Existing GPU and FPGA quantum simulators are largely optimized for full-state Schr\"odinger execution and therefore do not align well with HSF's path-centric workflow. This paper presents HSF-S, a compiler--accelerator co-designed framework for exact HSF-based quantum circuit emulation. HSF-S lowers input circuits to an HSF-compatible basis, formulates a rank-aware effective path-cost model, and applies dependency-preserving reordering together with discounted-gain SWAP insertion to suppress recurring cross-boundary interactions while preserving exact circuit semantics. 
A regression-free selector guarantees that the compiled circuit never increases effective path cost relative to the naive lowered baseline. 
We further design a dedicated HSF-S accelerator and execution flow, and integrate them into a stand-alone processor for efficient per-path dual-slice evaluation and final accumulation without materializing the full state vector. 
Across 56 benchmark circuits, HSF-S matches reference amplitudes to within floating-point precision, reduces effective path cost by up to 90.0\%, and substantially improves practical tractability, including representative timeout-to-sub-second reductions under a 1-hour budget. On the resulting compiled workloads, the HSF-S processor prototype delivers up to 4.34$\times$ additional speedup.
\end{abstract}


\maketitle

\vspace{-0.3em}
\section{Introduction}

Quantum computing has emerged as a prominent non-von Neumann computing paradigm with the potential to accelerate important optimization, simulation, and machine-learning workloads~\cite{Preskill:Quantum18, Cerezo:NatRevPhys21}. Yet current quantum hardware remains limited by qubit count, noise, connectivity, calibration stability, and programmability~\cite{Tannu:ASPLOS19, Murali:ASPLOS19, CarreraVazquez:Nature24,Choi:TC25}. Exact classical emulation therefore remains indispensable for validating algorithms, debugging compilation and mapping flows, generating ground-truth outputs, and enabling quantum--classical co-design~\cite{Li:SC21, Grurl:TCAD23, Choi2024, Choi:QIP26}.

Most existing quantum simulators and accelerators follow the Schr\"odinger model: they store the full $2^n$-dimensional state vector of an $n$-qubit circuit and apply gates by traversing it. GPUs and reconfigurable hardware have substantially improved this model through gate fusion, memory streaming, and data-layout optimization~\cite{Bayraktar:cuQuantum23,Wei:TC25,Choi:TC25,Liang:HPCA26}, but they cannot remove its fundamental exponential memory footprint. This limitation is especially severe on resource-constrained platforms, where on-chip storage, off-chip bandwidth, and implementation budgets are all limited.

Hybrid Schr\"odinger--Feynman (HSF) simulation offers a more attractive memory--path tradeoff by partitioning a circuit into two slices, evaluating each slice locally, and expanding only cross-bound- ary interactions into Feynman paths\PSC{~\cite{Arute:Nature19, Markov:DAC20}}. The remaining challenge is that runtime can still be dominated by the exponential growth of paths induced by cross-boundary two-qubit gates. In practice, a small set of boundary qubits often participates in repeated cross-slice interactions and disproportionately dominates the total path count. Practical HSF therefore requires not only fast execution, but also compile-time transformations that directly reduce recurring cross-boundary interactions.

Existing methods address only part of this problem. Static partitioning reduces boundary interactions but does not exploit gate order~\cite{Im:ASPDAC23}. Joint cutting is effective for contiguous cross blocks but less so for deeply interleaved circuits~\cite{Herzog:DAC25}. qsimh and GraFeyn reduce wall-clock latency through runtime parallelism without reducing path count itself~\cite{Markov:DAC20,Westrick:QCE24}. Prior SWAP insertion methods target routing or circuit-knitting objectives rather than the exact path-cost objective required by HSF~\PSC{\cite{Li:ASPLOS19, Tang:ASPLOS21,Ren:ICCAD24}}. At the same time, most existing GPU and FPGA accelerators are designed around full-state Schr\"odinger execution and do not match the execution structure of HSF, which consists of repeated per-path dual-slice evaluation followed by scalar accumulation\PSC{~\cite{Cicero:TQC25}}.

To address this gap, this paper presents HSF-S, a speed-optimized compilation and acceleration framework for exact HSF-based quantum circuit emulation. The key observation is that a cross-boundary SWAP, although expensive in isolation, can still reduce total HSF cost when it relocates a frequently interacting hub qubit and thereby removes multiple future cross-boundary interactions. HSF-S therefore treats SWAP as a path-cost-aware compile-time transformation rather than as a routing artifact. Starting from an HSF-compatible lowering pass, HSF-S applies two-stage dependency-preserving reordering and discounted-gain SWAP insertion to dynamically remap qubits across the partition boundary. The compiler emits both a transformed circuit and an index map for exact amplitude queries, and a selector guarantees that the final result is never worse than the naive lowered baseline in effective path cost. To execute the transformed workload efficiently, we further design a dedicated HSF-S accelerator and instruction/meta-data flow that exploit path independence for per-path dual-slice evaluation and final accumulation without materializing the full state vector. We integrate the design into a stand-alone RISC-V platform and validate it through FPGA prototyping and 14\,nm synthesis. Across 56 quantum-circuit benchmarks, HSF-S matches reference amplitudes up to floating-point precision, reduces effective path cost by up to 90.0\%, and achieves up to $2^{19}\!\times$ path reduction on QAOA-SBM workloads. On the same HSF-S-compiled workloads, the stand-alone processor prototype delivers an additional speedup of up to 4.34$\times$ over software-only qsimh execution on a 3.7\,GHz CPU, despite operating at only 200\,MHz.

\vspace{-0.2em}
\section{Background and Motivation}\label{sec:background}

\subsection{HSF and the Memory--Path Tradeoff}\label{sec:hsf}

Exact classical emulation commonly lies between two extremes. Schr\"odinger-style simulation stores the full state vector and is therefore limited by exponential memory growth, whereas Feynman-style simulation avoids materializing the full state but incurs exponential path growth~\cite{Aaronson:CCC17, Markov:DAC20}. Hybrid Schr\"odinger--Feynman (HSF) simulation combines the two by partitioning a circuit into two slices: intra-slice gates are evaluated locally in Schr\"odinger fashion, and only cross-boundary gates are expanded in Feynman fashion~\cite{Markov:DAC20}. For a balanced partition, this reduces peak storage from $O(2^n)$ to two slice-local states of size roughly $O(2^{n/2})$, but makes the total cost highly sensitive to the number and type of cross-boundary two-qubit gates.

For a cross-boundary gate $U_i$, HSF uses the operator Schmidt decomposition as follows\PSC{~\cite{Nielsen:PRA03}}:

\vspace*{-2em}
\begin{equation}\label{eq:schmidt}
    U_i = \sum_{m=1}^{r_i} \sigma_{i,m}(X_{i,m}\otimes Y_{i,m}),
\end{equation}
where $r_i$ is the Schmidt rank, $\sigma_{i,m}$ is the coefficient of the $m$-th term, and $X_{i,m}$ and $Y_{i,m}$ act only on slices $A$ and $B$, respectively. Any nontrivial controlled-$U$ gate used in practice can be written as
\begin{equation}\label{eq:controlled_u}
    CU = |0\rangle\langle 0| \otimes I + |1\rangle\langle 1| \otimes U,
\end{equation}
and therefore has rank at most two; for common HSF backends such as CX, CZ, CP, and $CR_z$, the rank is exactly two. By contrast, SWAP has rank four, the maximum possible for a two-qubit gate under a bipartition. Hence, not all cross-boundary gates incur the same HSF cost: a controlled gate doubles the number of paths, whereas a SWAP quadruples it.

If a circuit contains multiple cross-boundary gates, one Schmidt term must be chosen for each gate, so the total number of HSF paths is
\begin{equation}\label{eq:npaths}
    N_\text{paths} = \prod_i r_i.
\end{equation}
For a queried basis state $x=(x_A,x_B)$, the target amplitude is
\begin{equation}\label{eq:amplitude}
    \alpha(x) = \sum_{\mathbf{p}} \left(\prod_i \sigma_{i,p_i}\right)
    \langle x_A | C_A^{(\mathbf{p})} | 0_A \rangle\,
    \langle x_B | C_B^{(\mathbf{p})} | 0_B \rangle,
\end{equation}
where each path $\mathbf{p}$ fixes one Schmidt component for every cross-boundary gate. Distinct paths are independent until the final scalar accumulation, which is important both for runtime parallelism and for hardware acceleration.

Equivalently, the log-domain branching burden is $\sum_i \log_2 r_i$, so rank-2 controlled gates and rank-4 SWAP gates contribute $+1$ and $+2$, respectively. The resulting runtime can be written as
\begin{equation}\label{eq:thsf}
    T_\text{HSF} = 2^{\sum_i \log_2 r_i} \times W_\text{path},
\end{equation}
where $W_\text{path}$ is the per-path workload. In practice, the exponential term usually dominates, and the burden is often concentrated on a small number of boundary qubits that repeatedly participate in cross-slice interactions. This makes compile-time reduction of cross-boundary branching a natural optimization target.

A further practical point is that HSF backends such as \texttt{qsimh} do not generally execute arbitrary cross-boundary two-qubit unitaries: a cross-boundary gate is replaced by a Schmidt-decomposed pair of single-partition operations, and this decomposition is implemented only for a fixed set of gate kinds supported by the backend~\cite{Markov:DAC20, Burgholzer:QCE21, QuantumAI:qsim}. Gate lowering is therefore not merely a convenience pass. It is a correctness requirement that determines which cross-boundary operators are executable and what path cost they induce.

\vspace{-0.5em}

\subsection{Related Work and Design Gap}\label{sec:related}

Prior work reduces HSF cost from several angles. Static graph partitioning minimizes boundary interactions at the circuit-graph level, but it does not directly exploit temporal structure in the gate sequence~\cite{Im:ASPDAC23}. Joint cutting reduces the decomposition cost of contiguous cross-boundary blocks, but its benefit diminishes when cross interactions are deeply interleaved with local gates~\cite{Herzog:DAC25}. Runtime-parallel HSF frameworks such as qsimh and GraFeyn reduce wall-clock latency by distributing path evaluation, but they do not reduce path count itself~\cite{Markov:DAC20,Westrick:QCE24}. Meanwhile, prior SWAP insertion techniques mainly target hardware routing, nearest-neighbor execution, or circuit-knitting objectives rather than exact path reduction under an HSF-specific cost model~\cite{Tang:ASPLOS21,Ren:ICCAD24}.

The hardware side reveals a similar gap. Existing GPU and FPGA accelerators overwhelmingly target full-state Schr\"odinger simulation, organizing execution around one persistent state vector and optimizing its traversal~\cite{Bayraktar:cuQuantum23,Wei:TC25,Liang:HPCA26}. HSF, however, follows a different execution structure: repeated dual-slice Schr\"odinger evaluation across many independent paths followed by scalar accumulation. What is therefore missing is not hardware acceleration itself, but an integrated flow that simultaneously reduces HSF path growth at compile time and allows the compiled representation and its associated metadata to be consumed directly by an HSF-specific hardware execution model. Section~\ref{sec:pipeline} presents such a flow.


\vspace{-0.5em}
\section{HSF-S Preprocessing Pipeline}\label{sec:pipeline}

This section translates the path-complexity observations of Section~\ref{sec:hsf} into a concrete compilation flow. The key idea is to optimize HSF cost not by raw gate count, but by the cross-boundary branching burden induced under a fixed bipartition. HSF-S therefore combines HSF-compatible gate lowering, two-stage dependency-preserving reordering, discounted-gain SWAP insertion, and a re- gression-free selector into a unified preprocessing pipeline.

\vspace{-0.5em}
\subsection{Design Intuition: Why Cross SWAP Can Reduce HSF Cost}\label{sec:keyidea}

As discussed in Section~\ref{sec:hsf}, the Schmidt rank of a cross-boundary two-qubit gate determines the multiplicative growth of HSF paths. Under the lowered backend primitive set, cross-boundary controlled primitives have rank~2 and therefore contribute $+1$ in the log-domain, whereas cross-boundary SWAP has rank~4 and contributes $+2$. Based on this observation, we define the following \emph{effective path cost} as the primary compile-time objective:
\begin{equation}\label{eq:ceff}
    C_\text{eff} \triangleq \left|G_\text{cross}^{\neg\text{swap}}\right| + 2\left|G_\text{cross}^{\text{swap}}\right|.
\end{equation}
Here, $\left|G_\text{cross}^{\neg\text{swap}}\right|$ denotes the number of cross-boundary two-qubit gates other than SWAP, and $\left|G_\text{cross}^{\text{swap}}\right|$ denotes the number of cross-boundary SWAP gates. After lowering, Eq.~\eqref{eq:ceff} exactly matches the log-domain branching burden of the executable cross-boundary primitives. At the same time, $C_\text{eff}$ remains a compile-time surrogate cost rather than a direct predictor of wall-clock runtime, since runtime also depends on the per-path workload and hardware execution efficiency.

The key observation is that SWAP is not always harmful in HSF. Consider inserting a SWAP between a hub qubit $q_h$, which repeatedly participates in cross-boundary interactions, and a candidate qubit $q_c$ in the opposite slice. Within a lookahead window, let $k_\text{resolved}$ be the number of gates that were previously cross-boundary but become intra-slice after the SWAP, and let $k_\text{created}$ be the number of gates that become newly cross-boundary. The resulting local change in effective path cost is
\begin{equation}\label{eq:delta_ceff}
    \Delta C_\text{eff} = 2 - k_\text{resolved} + k_\text{created},
\end{equation}
where the constant $2$ is the rank-based contribution of the inserted cross SWAP itself. Therefore,
$
    k_\text{resolved} - k_\text{created} > 2
$
is a sufficient condition for the SWAP to reduce $C_\text{eff}$. This interpretation turns SWAP from a routing artifact into a \emph{path-cost-aware compile-time transformation}.

\begin{figure}[t]
\vskip -6pt
    \centering
    \includegraphics[width=\columnwidth]{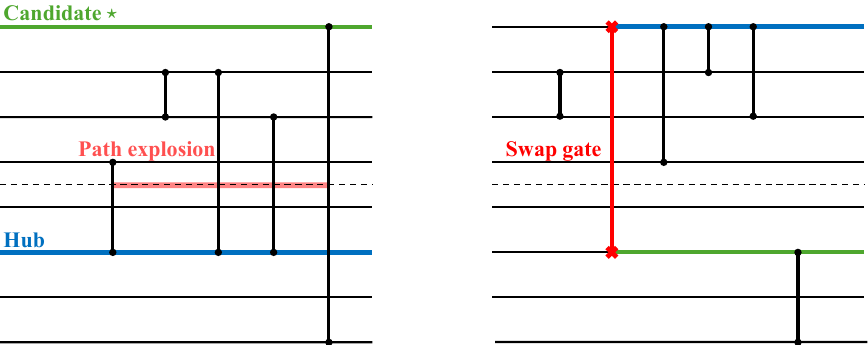}
    \vskip -4pt
    \caption{Key intuition of HSF-S. Under a fixed partition, a hub qubit repeatedly induces cross-boundary interactions. A cross SWAP can relocate the hub and convert downstream cross-boundary interactions into intra-slice gates, reducing HSF path cost.}
    \label{fig:swap_concept}
\end{figure}

\refFigure{fig:swap_concept} illustrates the idea. In the left circuit, the hub qubit participates in several cross-boundary interactions and dominates the local branching burden. In the right circuit, one inserted cross SWAP relocates that hub, after which the downstream interactions in the highlighted region become intra-slice. In the illustrated example, the SWAP resolves four cross-boundary interactions and creates none, yielding
$
\Delta C_\text{eff} = 2 - 4 + 0 = -2,
$
so the effective path cost in that region decreases from $+4$ to $+2$. The important point is that the cost of the SWAP itself is not evaluated in isolation; it is evaluated against the downstream cross-boundary burden that it removes.

\vspace{-0.5em}
\subsection{Pipeline Overview}\label{sec:overview}

\begin{figure}[t]
\vskip -6pt
    \centering
    \includegraphics[width=0.85\columnwidth]{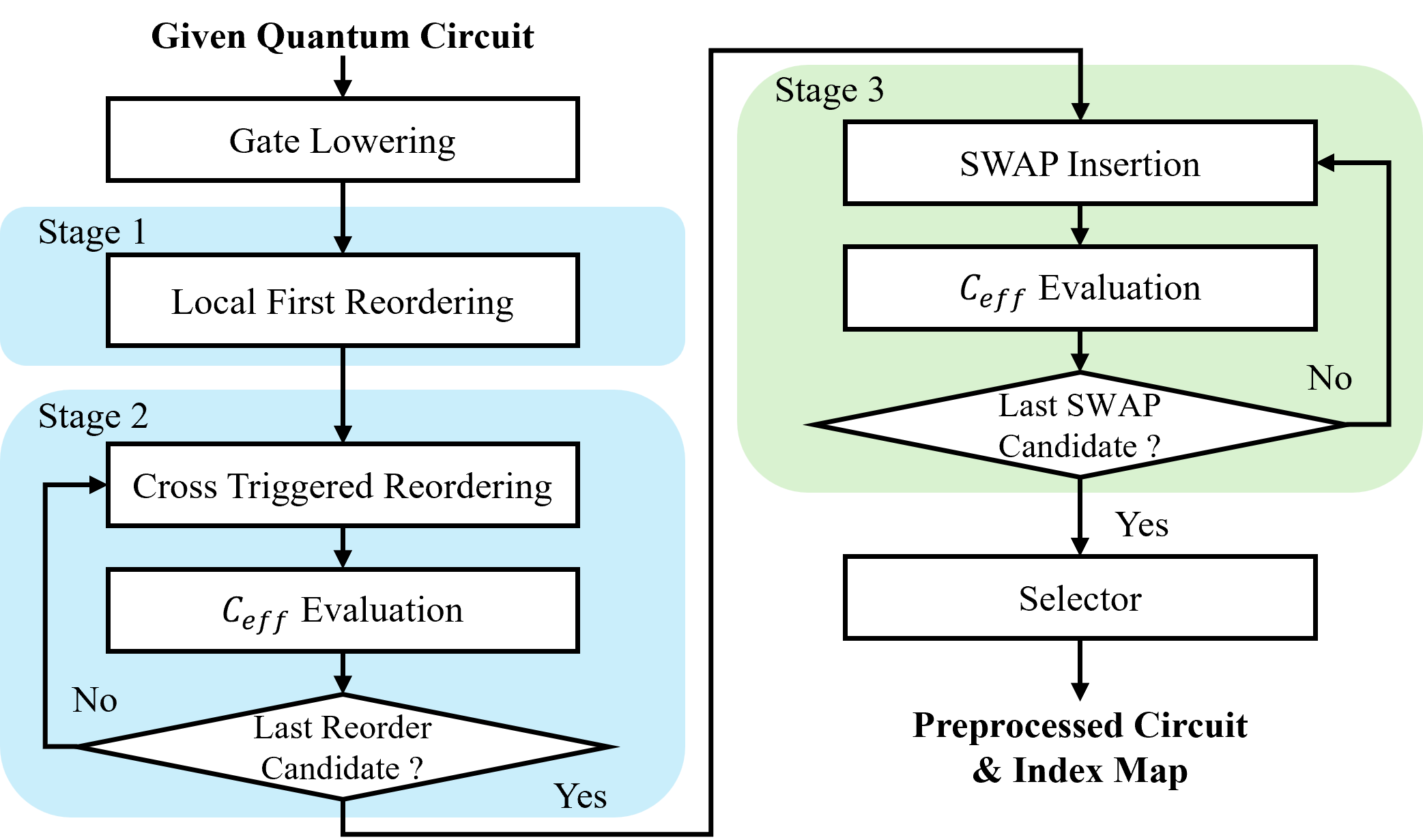}
    \vskip -4pt
    \caption{Overview of the HSF-S preprocessing flow. After gate lowering, two reordering stages expose profitable SWAP opportunities, Stage~3 inserts cross SWAPs, and the Selector returns the best compiled circuit together with metadata.}
    \label{fig:pipeline}
    \vskip -4pt
\end{figure}

As shown in \refFigure{fig:pipeline}, HSF-S consists of one lowering pre-pass, two reordering stages, one SWAP insertion stage, and a final Selector. The pipeline is not a single-pass heuristic. Instead, the reordering and SWAP-insertion stages sweep small parameter sets, generate multiple candidates, and let the Selector choose the best result.

We assume a fixed partition descriptor $b$ that assigns each physical qubit position to slice $A$ or $B$. For any gate or node $n$, let $\mathrm{supp}(n)$ denote its operand qubits. For any ordered gate sequence $\mathcal{O}$, let $\mathcal{O}[a:b]$ denote the subsequence from position $a$ to $b$, inclusive. Given an input circuit $\mathcal{C}$, the lowering pre-pass first normalizes the circuit into the primitive set
\[
\mathcal{B}_\text{lower} = \{H, X, Y, Z, SX, P, R_x, R_y, R_z, CX, CP\}.
\]
Here, $SX$ denotes the square-root-of-$X$ gate, and $P$ and $CP$ denote the single-qubit phase gate and the controlled-phase gate, respectively.
\PSC{Since $\mathcal{B}_\text{lower}$ supports arbitrary single-qubit operations and an entangling two-qubit primitive, arbitrary unitary circuits can be exactly lowered into this basis~\cite{qcqibook,Shende:ASPDAC05}.}
The lowered circuit before any further optimization is denoted by $\mathcal{C}_{\mathrm{naive}}$. Stage~3 may then insert an additional \emph{cross SWAP primitive}, so the executable primitive set becomes
$
\mathcal{B}_\text{exec} = \mathcal{B}_\text{lower} \cup \{\mathrm{SWAP}\}.
$

The key effect of lowering is that every executable cross-boundary two-qubit gate has a fixed Schmidt rank: either rank~2 for controlled primitives or rank~4 for SWAP. As a result, $C_\text{eff}$ in Eq.~\eqref{eq:ceff} can be computed exactly at compile time without invoking the simulator. This is the basis on which HSF-S directly optimizes path branching.


For any circuit or local window $X$, let $G_\text{cross}^{2q}(X)$ denote the set of cross-boundary two-qubit gates contained in $X$ under the current partition descriptor and mapping. When the argument is omitted, $G_\text{cross}^{2q}$ refers to the corresponding set in the current circuit context.

After lowering, Stage~1 performs circuit-wide local-first reordering to temporally cluster cross-boundary gates, while Stage~2 performs trigger-centered local reordering around actual cross-bound- ary positions. These two stages do not change the partition and do not insert any new gates; instead, they use only the temporal freedom permitted by dependencies to expose profitable SWAP opportunities for Stage~3. Stage~3 then tracks the current qubit mapping through a shadow state and inserts SWAPs only when the discounted gain is non-negative.

The final Selector compares all candidate circuits using the lexicographic key
$
    \kappa = \left(C_\text{eff},\; |G_\text{cross}^{2q}|,\; |S|,\; |G|\right),
$
where $G_\text{cross}^{2q}$ denotes the set of cross-boundary two-qubit gates in the candidate circuit, $S$ denotes the set of inserted SWAP gates, and $G$ denotes the full gate sequence. Thus, the primary objective is to minimize $C_\text{eff}$; remaining cross-boundary two-qubit gates, inserted SWAP count, and total gate count are used only as tie-breakers.

Importantly, the Selector always includes $\mathcal{C}_{\mathrm{naive}}$ in the candidate set. 
Therefore, if $\mathcal{C}^{*}$ denotes the selected result,
$
    C_\text{eff}(\mathcal{C}^{*}) \le C_\text{eff}(\mathcal{C}_{\mathrm{naive}}).
$
That is, HSF-S is heuristic, but it is never allowed to return a circuit whose effective path cost is worse than that of the naive lowered baseline.

The compiler outputs the optimized circuit $\mathcal{C}^{*}$, the final logical-to-physical map $\mathrm{L2P}^*$, and the cut-rank metadata extracted from the executable cross-boundary gate sequence. These artifacts are consumed directly by the hardware flow in Section~4.

\vspace{-0.5em}
\subsection{Reordering Stages}\label{sec:reorder}

The purpose of reordering is not merely to ``clean up'' the schedule. Under a fixed partition descriptor $b$, whether a gate is cross-boundary or intra-slice is determined by operand location, so reordering alone does not directly change $C_\text{eff}$. Its role is instead to expose profitable SWAP opportunities by making cross-boundary interactions temporally denser and by bringing repeated hub-centered patterns into shorter local windows. In other words, Stage~1 and Stage~2 are designed to improve the decision quality of Stage~3.

\subsubsection{Stage~1: Local-First Reordering}

Stage~1 performs a greedy topological sort over the commutation-aware dependency DAG of the entire circuit. For every ready node $n$, it evaluates
\begin{equation}\label{eq:score}
    \mathrm{score}(n) =
    -W_\mathrm{cross}\cdot \mathbf{1}[\mathrm{is\_cross\_2q}(n)]
    -W_\mathrm{stable}\cdot \mathrm{pos}(n),
\end{equation}
where $\mathbf{1}[\cdot]$ is the indicator function, $W_\mathrm{cross}$ is a large penalty that delays cross-boundary two-qubit gates, and $W_\mathrm{stable}$ weakly preserves the original relative order through the position index $\mathrm{pos}(n)$. In practice, we choose $W_\mathrm{cross} \gg W_\mathrm{stable}$ so that any ready intra-slice gate is preferred over a ready cross-boundary two-qubit gate.

As a result, Stage~1 schedules local gates as early as possible while pushing cross-boundary interactions toward each other in time. This stage neither changes the partition nor inserts new gates, so it does not reduce $C_\text{eff}$ by itself. Its value lies in increasing the density of relevant cross interactions inside subsequent local windows. Correctness follows directly from dependency preservation: any topological order of the same DAG implements the same unitary.

\subsubsection{Stage~2: Cross-Triggered Local Reordering}

Stage~2 no longer reorders the entire circuit. Instead, it scans the Stage~1 result and opens a local reorder window only when it encounters an actual cross-boundary two-qubit trigger. This stage sweeps two parameters: the local window length $L_\text{reorder}$ and a heuristic profile $\rho \in \{\mathrm{hub}, \mathrm{chain}\}$. To avoid running the full Stage~3 parameter sweep for every reorder candidate, each $\bigl(L_\text{reorder},\rho\bigr)$ candidate is evaluated with a fixed low-cost probe configuration of Stage~3, and only the candidate with the best resulting $\kappa$ is forwarded as $\mathcal{O}_2$.

Algorithm~\ref{alg:cross} shows the procedure for a single parameter pair $\bigl(L_\text{reorder},\rho\bigr)$. Starting from the Stage~1 order $\mathcal{O}_1$, the algorithm maintains a current working order $\mathcal{O}_r$ and scans it from left to right. If the current gate is not cross-boundary under $b$, it advances without modification. Otherwise, it opens a local window
$
\mathcal{W}_i = \mathcal{O}_r[i:\min(i+L_\text{reorder}-1,N)]
$
and builds a dependency DAG $D_i=(V_i,E_i)$ only on that window. The search therefore focuses on the region where the current trigger actually contributes to path burden, rather than on the entire circuit.

To guide scheduling inside the window, Stage~2 first computes a window-local hub score
\begin{equation}\label{eq:hubscore}
    h_i(q) =
    \sum_{m\in G_\text{cross}^{2q}(\mathcal{W}_i)}
    \mathbf{1}\!\left[q \in \mathrm{supp}(m)\right],
\end{equation}
which measures how frequently each qubit participates in cross-boundary two-qubit interactions inside $\mathcal{W}_i$. Let $\nu$ denote the current reference cross-boundary gate. The function
\begin{equation}\label{eq:hub_reorder}
    \mathrm{Hub}_i(\nu) =
    \arg\max_{q \in \mathrm{supp}(\nu)}
    \bigl(h_i(q),\; {-}q\bigr)
\end{equation}
returns the operand of $\nu$ with the higher local hub score, with the smaller qubit index used as a deterministic tie-breaker. Initially, $\nu$ is the trigger gate $g^\star$, and the active hub variable is initialized as $q_h \leftarrow \mathrm{Hub}_i(g^\star)$. Whenever Stage~2 appends a cross-boundary two-qubit gate to the reordered window, the active hub is refreshed as $q_h \leftarrow \mathrm{Hub}_i(n^\star)$.


\begin{algorithm}[t]
\footnotesize
\caption{Cross-triggered reordering.}\label{alg:cross}
\begin{algorithmic}[1]
\STATE \textbf{Input:} $\mathcal{O}_1, b, L_\mathrm{reorder}, \rho$
\STATE \textbf{Output:} $\mathcal{O}_r$
\STATE \textbf{function} \textsc{Cross\_Triggered\_Reorder}($\mathcal{O}_1, b, L_\mathrm{reorder}, \rho$)
\STATE \quad $\mathcal{O}_r \leftarrow \mathcal{O}_1$;\; $N \leftarrow |\mathcal{O}_r|$;\; $i \leftarrow 1$
\STATE \quad \textbf{while} $i \le N$ \textbf{do}
\STATE \qquad \textbf{if} $\mathcal{O}_r[i]$ is not cross-boundary under $b$ \textbf{then} $i \leftarrow i+1$; \textbf{continue} \textbf{end if}
\STATE \qquad $g^\star \leftarrow \mathcal{O}_r[i]$;\; $\mathcal{W}_i \leftarrow \mathcal{O}_r[i:\min(i+L_\mathrm{reorder}-1,N)]$
\STATE \qquad build $D_i=(V_i,E_i)$ on $\mathcal{W}_i$;\; initialize $q_h \leftarrow \mathrm{Hub}_i(g^\star)$
\STATE \qquad $\widetilde{\mathcal{W}}_i \leftarrow [\,]$;\; $t \leftarrow 0$;\; $S_t \leftarrow \varnothing$
\STATE \qquad \textbf{while} $|S_t| < |V_i|$ \textbf{do}
\STATE \qquad \quad $R_t \leftarrow \{\,n \in V_i \setminus S_t \mid \mathrm{pred}(n)\subseteq S_t\,\}$
\STATE \qquad \quad $n^\star \leftarrow \arg\max_{n \in R_t}\pi_\rho(n;t)$;\; append $n^\star$ to $\widetilde{\mathcal{W}}_i$
\STATE \qquad \quad \textbf{if} $n^\star \in G_\text{cross}^{2q}(\mathcal{W}_i)$ \textbf{then} $q_h \leftarrow \mathrm{Hub}_i(n^\star)$ \textbf{end if}
\STATE \qquad \quad $S_{t+1} \leftarrow S_t \cup \{n^\star\}$;\; $t \leftarrow t+1$
\STATE \qquad \textbf{end while}
\STATE \qquad \textbf{if} $\widetilde{\mathcal{W}}_i \neq \mathcal{W}_i$ \textbf{then}
\STATE \qquad \quad replace $\mathcal{W}_i$ in $\mathcal{O}_r$ with $\widetilde{\mathcal{W}}_i$;\; $i \leftarrow i + |\mathcal{W}_i|$
\STATE \qquad \textbf{else} $i \leftarrow i+1$ \textbf{end if}
\STATE \quad \textbf{end while}
\STATE \quad \textbf{return} $\mathcal{O}_r$
\STATE \textbf{end function}
\end{algorithmic}
\end{algorithm}

Stage~2 then repeatedly selects one node from the ready set
\[
R_t = \{\,n \in V_i \setminus S_t \mid \mathrm{pred}(n)\subseteq S_t\,\}
\]
and appends it to the new local order $\widetilde{\mathcal{W}}_i$. The choice is made by maximizing the following lexicographic priority:
\begin{equation}\nonumber
\pi_\rho(n;t)=
\begin{cases}
\begin{aligned}[t]
\bigl(&\phi_\mathrm{cross}(n),\; \phi_\mathrm{hub}(n;t),\; c_\mathrm{hub}(n),\\
      &\phi_\mathrm{share}(n;t),\; \phi_\mathrm{diag}(n),\; -\mathrm{pos}(n)\bigr)
\end{aligned}
& \rho=\mathrm{hub}, \\[4pt]
\begin{aligned}[t]
\bigl(&\phi_\mathrm{cross}(n),\; \phi_\mathrm{pair}(n;t),\; \phi_\mathrm{hub}(n;t),\\
      &\phi_\mathrm{share}(n;t),\; c_\mathrm{hub}(n),\; \phi_\mathrm{diag}(n),\; -\mathrm{pos}(n)\bigr)
\end{aligned}
& \rho=\mathrm{chain}.
\end{cases}
\end{equation}
Here, $\phi_\mathrm{cross}(n)$ indicates whether $n$ is cross-boundary, $\phi_\mathrm{hub}(n;t)$ indicates whether $n$ contains the active hub $q_h$, $c_\mathrm{hub}(n)$ is the sum of local hub scores over the qubits in $\mathrm{supp}(n)$, $\phi_\mathrm{pair}(n;t)$ indicates whether $n$ uses the same two-qubit pair as the previously scheduled node, $\phi_\mathrm{share}(n;t)$ indicates whether $n$ shares at least one qubit with the previously scheduled node, $\phi_\mathrm{diag}(n)$ indicates whether $n$ is diagonal in the computational basis, and $\mathrm{pos}(n)$ is the original position of $n$ inside $\mathcal{W}_i$. 
At $t=0$, we set $\phi_\mathrm{pair}(n;0)=0$ and $\phi_\mathrm{share}(n;0)=0$. In this work, diagonal gates include gates whose unitary is diagonal in the computational basis, such as $Z$, $P$, $R_z$, and $CP$.


Intuitively, the hub profile favors windows in which one heavily loaded qubit repeatedly interacts with multiple partners, whereas the chain profile favors repeated pairwise continuity.
The use of lexicographic order, rather than a weighted sum, prevents accidental cancellation between features and keeps the scheduling rule\linebreak[4] 

\noindent interpretable.

After the local schedule is complete, Stage~2 replaces the original window only if $\widetilde{\mathcal{W}}_i \neq \mathcal{W}_i$; otherwise, it simply advances to the next position. Because all transformations remain within the topological freedom of the local DAG, correctness is preserved.

\vspace{-0.35em}
\subsection{{Stage~3}: Discounted-Gain SWAP Insertion}\label{sec:swap}

Stage~3 scans the Stage~2 output $\mathcal{O}_2$ once from left to right while explicitly tracking the current mapping state. We implement this as a \emph{shadow mapping}: rather than rebuilding a dependency graph, Stage~3 maintains a single logical-to-physical map $\mathrm{L2P}$ and reinterprets each subsequent gate under the updated mapping whenever a SWAP is inserted. In this way, SWAP changes the meaning of future gate locations without changing the logical gate order itself.

The objective of Stage~3 is not to avoid the current trigger in isolation, but to reduce the \emph{accumulated} cross-boundary burden over the upcoming gate sequence. This structure is analogous to classical bi-partitioning heuristics such as Kernighan--Lin~\cite{Kernighan:BSTJ70}: logical qubits correspond to vertices, two-qubit gates in the lookahead window correspond to edges, and a cross SWAP corresponds to exchanging a vertex across the partition. However, our setting differs in three important ways: the exchange cost is nonzero ($+2$ in $C_\text{eff}$), decisions affect only future gates because the gate sequence has a time axis, and the algorithm runs as a greedy single forward pass rather than an iterative convergence process.

To control how aggressively Stage~3 looks ahead, HSF-S sweeps the lookahead length $L_\text{swap}$ together with a half-life parameter $H$ that determines the discount factor $\gamma = 2^{-1/H}$. For clarity, Algorithm~\ref{alg:swap} is written directly in terms of $\gamma$. In the actual sweep, each $H$ instantiates a corresponding $\gamma$, and we additionally evaluate an undiscounted variant with $\gamma=1$.

\begin{algorithm}[t]
\footnotesize
\caption{SWAP insertion.}\label{alg:swap}
\begin{algorithmic}[1]
\STATE \textbf{Input:} $\mathcal{O}_2, b, L_\text{swap}, \gamma$
\STATE \textbf{Output:} $\mathcal{C}_{\mathrm{out}}, \mathrm{L2P}$
\STATE \textbf{function} \textsc{Swap\_Insert}($\mathcal{O}_2, b, L_\text{swap}, \gamma$)
\STATE \quad $\mathcal{C}_{\mathrm{out}} \leftarrow [\,]$;\; $\mathrm{L2P}[\ell] \leftarrow \ell\;\; \forall \ell$
\STATE \quad \textbf{for} $t = 1$ to $|\mathcal{O}_2|$ \textbf{do}
\STATE \qquad map $\mathcal{O}_2[t]$ by the current $\mathrm{L2P}$
\STATE \qquad \textbf{if} the mapped gate is not cross-boundary under $b$ \textbf{then} emit it to $\mathcal{C}_{\mathrm{out}}$; 
\STATE \qquad \qquad \textbf{continue} 
\STATE \qquad \textbf{end if}
\STATE \qquad form the lookahead window of length $L_\text{swap}$
\STATE \qquad compute $w_\text{cut}(q)$ and $w_\text{act}(q)$ in the window
\STATE \qquad $q_h \leftarrow \arg\max_q \bigl(w_\text{cut}(q),\; {-}w_\text{act}(q),\; {-}q\bigr)$
\STATE \qquad choose $q_c^\star$ in the opposite slice by maximizing $g_\gamma(q_h,q_c)$
\STATE \qquad \textbf{if} $g_\gamma(q_h,q_c^\star) \ge 0$ \textbf{then}
\STATE \qquad \quad emit $\mathrm{SWAP}(\mathrm{L2P}[q_h],\mathrm{L2P}[q_c^\star])$ to $\mathcal{C}_{\mathrm{out}}$;\; $\mathrm{L2P}[q_h] \leftrightarrow \mathrm{L2P}[q_c^\star]$
\STATE \qquad \textbf{end if}
\STATE \qquad remap $\mathcal{O}_2[t]$ using the updated $\mathrm{L2P}$;\; emit it to $\mathcal{C}_{\mathrm{out}}$
\STATE \quad \textbf{end for}
\STATE \quad \textbf{return} $\mathcal{C}_{\mathrm{out}},\;\mathrm{L2P}$
\STATE \textbf{end function}
\end{algorithmic}
\end{algorithm}


Initially, the output circuit $\mathcal{C}_{\mathrm{out}}$ is empty and $\mathrm{L2P}$ is the identity map. The main loop remaps each gate under the current $\mathrm{L2P}$. If the mapped gate is intra-slice, it does not contribute to cross-boundary branching and is emitted immediately. Otherwise, when the mapped gate is a cross-boundary two-qubit trigger, Stage~3 opens a lookahead window of length $L_\text{swap}$ that \emph{includes the current trigger}. Let $\mathcal{W}_t^{2q}$ denote the set of mapped two-qubit gates inside that window, and let $k_e$ denote the relative distance of gate $e$ from the current trigger, with $k_e=0$ for the trigger itself. Since farther gates are more likely to be affected by future remapping decisions, Stage~3 discounts their contribution using $\gamma^{k_e}$.

Specifically, for each qubit $q$, we compute the discounted cross involvement as follows:
\newpage

\vspace*{-2.4em}
\begin{equation}\label{eq:wcut}
\setlength{\abovedisplayskip}{1pt}
\setlength{\belowdisplayskip}{1pt}
    w_\text{cut}(q) =
    \sum_{e \in \mathcal{W}_t^{2q}}
    \mathbf{1}\!\left[e \text{ is cross} \wedge q \in \mathrm{supp}(e)\right]\gamma^{k_e}
\end{equation}
and the discounted total activity
{
\setlength{\abovedisplayskip}{2pt}
\setlength{\belowdisplayskip}{1pt}
\begin{equation}\label{eq:wact}
    w_\text{act}(q) =
    \sum_{e \in \mathcal{W}_t^{2q}}
    \mathbf{1}\!\left[q \in \mathrm{supp}(e)\right]\gamma^{k_e}.
\end{equation}
}
Here, $w_\text{cut}(q)$ measures how much cross-boundary burden a qubit currently carries, while $w_\text{act}(q)$ estimates how disruptive it may be to move that qubit. The SWAP hub is then chosen as
\begin{equation}\label{eq:hub_swap}
    q_h \leftarrow \arg\max_q \bigl(w_\text{cut}(q),\; {-}w_\text{act}(q),\; {-}q\bigr),
\end{equation}
that is, the qubit with the highest discounted cross burden and, among ties, the smallest discounted total activity. The final tie-breaker $-q$ makes the decision deterministic.


Given the hub qubit, Stage~3 evaluates candidate qubits $q_c$ only in the opposite slice. For each candidate, it computes how the cross-boundary status of each gate would change under a hypothetical SWAP between $q_h$ and $q_c$ in the current shadow mapping. Let
\[
c_\mathrm{before}(e),\; c_\mathrm{after}(e; q_h, q_c) \in \{0,1\}
\]
denote the cross-boundary status of gate $e$ before and after the hypothetical SWAP, respectively. The discounted gain is then
\begin{equation}\label{eq:gain}
\setlength{\abovedisplayskip}{3pt}
\setlength{\belowdisplayskip}{2pt}
    g_\gamma(q_h, q_c) =
    \sum_{e \in \mathcal{W}_t^{2q}}
    \Bigl[
        c_\mathrm{before}(e) - c_\mathrm{after}(e; q_h, q_c)
    \Bigr]\gamma^{k_e} - 2.
\end{equation}
The constant $2$ is the contribution of the inserted cross SWAP itself to $C_\text{eff}$. The chosen candidate $q_c^\star$ is the one that maximizes $g_\gamma(q_h,q_c)$, and the actual commit condition is
\begin{equation}\label{eq:commit}
\setlength{\abovedisplayskip}{2pt}
\setlength{\belowdisplayskip}{2pt}
    g_\gamma(q_h, q_c^\star) \ge 0.
\end{equation}
In the undiscounted case $\gamma=1$, this reduces to the intuitive condition that the number of resolved cross-boundary interactions must exceed the number of created ones by at least two. Under discounting, Eq.~\eqref{eq:commit} is the corresponding weighted generalization.

If the gain condition is satisfied, Stage~3 inserts a SWAP between the current physical locations of $q_h$ and $q_c^\star$, and immediately updates the shadow mapping by exchanging $\mathrm{L2P}[q_h]$ and $\mathrm{L2P}[q_c^\star]$. The current trigger gate is then remapped under the new $\mathrm{L2P}$ and emitted. Thus, the inserted SWAP does not merely add a gate; it changes how all subsequent gates are interpreted.

After the full scan, Stage~3 returns $(\mathcal{C}_{\mathrm{out}},\mathrm{L2P})$. \PSC{Using the final index map $\mathrm{L2P}$, a logical basis query $x_\ell$ can be converted into the physical ordering expected by the compiled circuit as
\begin{equation}\label{eq:l2p}
    x_p[\mathrm{L2P}[\ell]] = x_\ell[\ell], \qquad \forall \ell \in \{0,\ldots,n-1\}.
\end{equation}}
Therefore, query-driven HSF execution does not require an explicit permutation of the full state; it is sufficient to transform the query index itself. This index map is also consumed directly as hardware metadata in Section~4.

In the implementation, Stage~3 does not exhaustively evaluate all qubits in the opposite slice. Instead, it first ranks them by increasing $w_\text{act}$ and evaluates only the first $P$ candidates, which reduces the search overhead while preserving good empirical quality. After this prefiltering, \LKW{the cost of one Stage~3 run is approximately $O(|\mathcal{O}_2|  \cdot P \cdot L_\text{swap})$.}

Finally, Stage~3 is a greedy single forward pass and therefore does not guarantee a global optimum. In practice, however, three mechanisms compensate for this limitation. First, Stage~1 and Stage~2 improve the quality of local lookahead decisions by making relevant interactions denser in time. Second, the sweep over $L_\text{swap}$ and $\gamma$ explores both short-horizon and long-horizon decision regimes. Third, the Selector always retains the naive lowered baseline, so the final result remains regression-free. Across the full pipeline, Stage~3 is repeated for all candidate $(L_\text{swap},\gamma)$ settings, and the Selector returns the compiled circuit and metadata with the smallest $\kappa$.


\section{HSF-S Accelerator}

\begin{figure*}[t]
\vskip -6pt
\centering
\includegraphics[width=\linewidth]{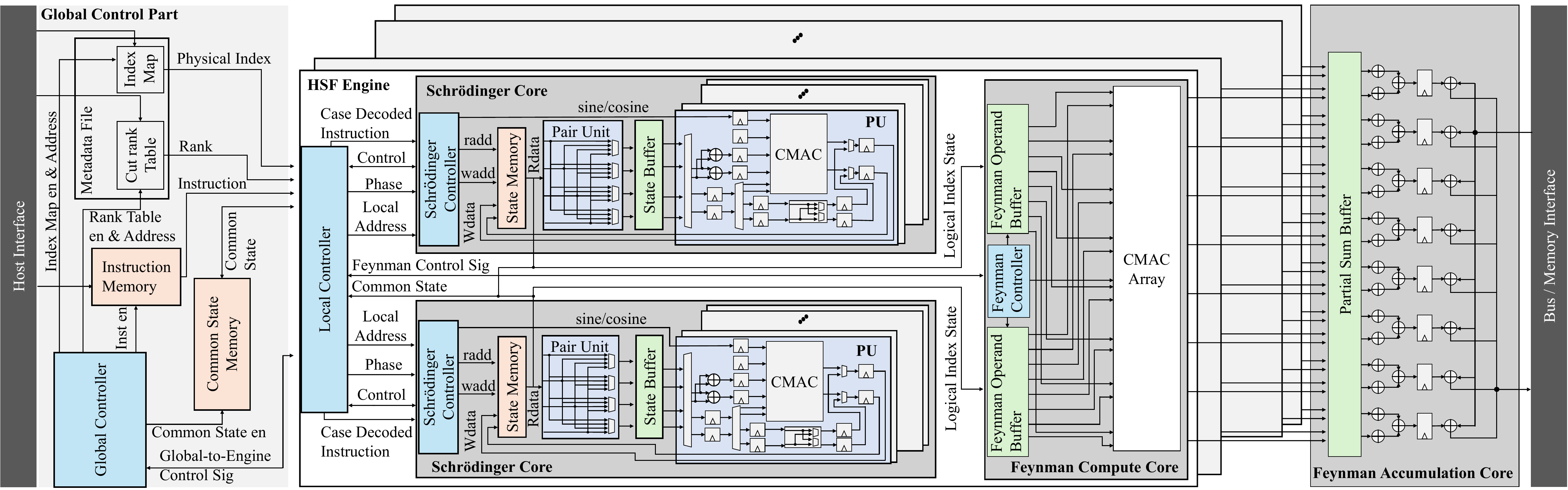}
\vskip -4pt
\caption{Overview of the proposed HSF-S accelerator. The host loads the compiled instruction stream together with cut-rank table and index map. A global controller broadcasts instructions to parallel HSF engines. Each engine contains two slice-local Schr\"odinger cores and a Feynman Compute Core, while a common-state memory and a shared Feynman Accumulation Core support prefix reuse and cross-path accumulation.}
\label{fig:architecture}
\vskip -6pt
\end{figure*}

\subsection{Architecture Overview and Execution Flow}\label{sec:hsf_arch}
\vspace*{0.1em}

The proposed accelerator is designed to execute not only HSF-S-compiled circuits, but also generic HSF workloads. The hardware consumes the artifacts produced by the preprocessing flow in Section~\ref{sec:pipeline}: the compiled instruction stream, the cut-rank metadata for cross-boundary gates, and the final index map. When no compile-time qubit remapping is applied, the index map reduces to the identity permutation, so the same hardware can execute conventional HSF workloads without architectural modification. At the system level, the accelerator is integrated into a stand-alone Rocket-core-based RISC-V platform, where the host processor loads instructions and metadata, launches execution, and reads back the accumulated amplitudes from memory.

As shown in \refFigure{fig:architecture}, the top-level accelerator consists of a Global Controller, Instruction Memory, Metadata File, \MDJ{Common State Memory,} multiple parallel HSF Engines, a shared Feynman Accumulation Core, and the bus/memory interface. Path-level parallelism is exposed by instantiating multiple HSF engines, each of which evaluates a different path case. 
For each execution batch, the Global Controller issues a base path index, and the Local Controller of each HSF engine derives its current path identifier by adding the engine offset to this base index. When the total number of paths exceeds the number of engines, the path space is processed in multiple batches. The Global Controller is responsible for issuing the common instruction stream, coordinating metadata access, and synchronizing path execution across engines.

Each HSF engine contains a Local Controller, two slice-local Schr\"odinger Cores corresponding to slices $A$ and $B$, and a Feynman Compute Core. The Local Controller decodes the incoming instruction stream and determines whether the current instruction is purely local or cross-boundary. For local gates, the same opcode is broadcast to all engines, and both slice-local Schr\"odinger Cores execute the corresponding local update. 
For a cross-boundary gate, the controller uses the cut-rank-table entry embedded in the instruction together with the current path identifier to determine the path-specific local micro-operations for the two slices. 
More precisely, the cut-rank table provides the branch cardinality of the gate, while the gate opcode and the current path identifier determine the corresponding pair of slice-local operators according to the backend's fixed decomposition rule. In this way, one cross-boundary instruction in the compiled circuit is translated into the path-specific local operations required by HSF.

\begin{figure}[t]
\vskip -3pt
\centering
\includegraphics[width=0.95\columnwidth]{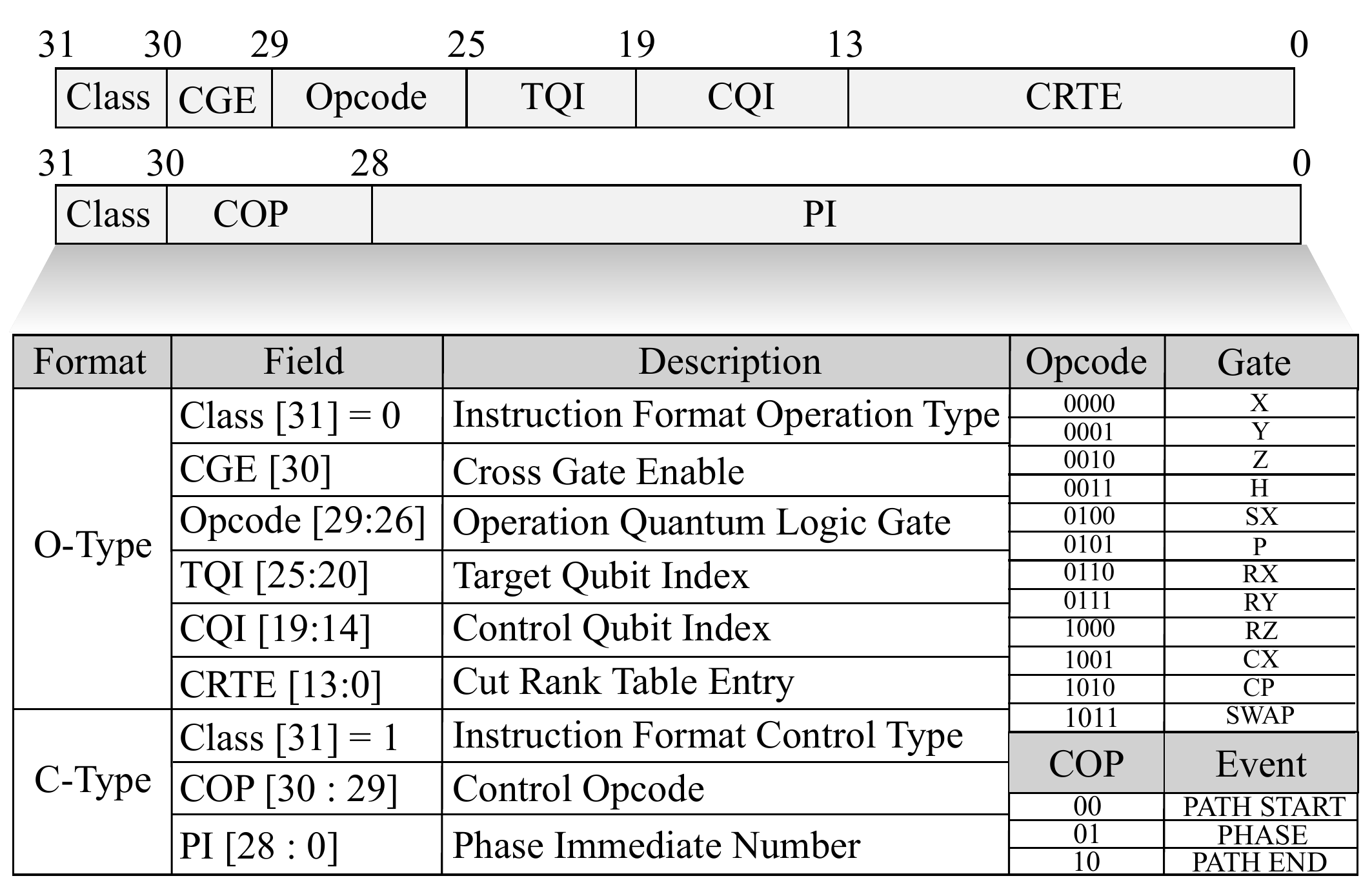}
\vskip -10pt
\caption{Instruction format of the proposed HSF-S accelerator.}
\label{fig:bitfield}
\vskip -4pt
\end{figure}

Each Schr\"odinger Core consists of a Schr\"odinger Controller, State Memory, a Pair Unit, a State Buffer, and a Processing Unit (PU). The Pair Unit fetches the amplitudes associated with the target qubit and presents them to the PU in the pairwise form used by state-vector updates. The PU supports the lowered gate basis of Section~\ref{sec:pipeline} together with a small set of internal micro-operations required by cross-boundary decomposition. For phase and rotation operations, the required coefficients are delivered through the control path and the local sine/cosine datapath, \MDJ{where the sine/cosine values are generated by the Schrödinger controller.} 
The updated amplitudes are then written back to the local State Memory.

A key architectural observation is that all HSF engines receive the same instruction stream, so their slice states remain identical until the first cross-boundary gate is reached. Re-executing this common prefix for every path would therefore waste time, especially when the first cross-boundary gate appears late in the circuit. To avoid this overhead, the accelerator checkpoints the slice states immediately before the first cross-boundary gate and reuses this common state as the starting point of subsequent path executions. In the current design, only the common prefix before the first cross-boundary gate is checkpointed, which captures the shared redundant work while keeping the control logic simple. The first path executes the full prefix and writes the checkpointed slice states into the Common State Memory, whereas later paths resume directly from this checkpoint and execute only the divergent suffix. This optimization reduces redundant work without changing the semantics of HSF execution.

Once the two slice-local Schr\"odinger evaluations for the current path are complete, the engine extracts only the amplitudes needed for the requested full-circuit basis indices. When HSF-S preprocessing is used, the logical query index is first translated into the physical ordering through the compiler-generated index map. The Global Controller distributes the translated physical index to all engines, and each engine splits it into the slice-local addresses required by the two Schr\"odinger Cores. The Feynman Compute Core then reads the corresponding slice amplitudes, stores them in the Feynman Operand Buffer, computes their complex product in the CMAC Array, 
and applies the path-dependent scaling factor to reconstruct the partial contribution of the full circuit.

The partial outputs of all HSF engines are forwarded to the shared Feynman Accumulation Core, which performs the remaining accumulation across engines and across paths. Since the number of output amplitudes may exceed on-chip storage, the backend uses buffered streaming to merge the current partial sums with previously accumulated values stored in main memory. The Feynman Accumulation Core performs the accumulation, and the final amplitudes are written back through the Memory Interface and later read by the host processor. This organization follows the natural execution structure of HSF: slice-local Schr\"odinger evaluation, cross-slice reconstruction, and cross-path accumulation.

\subsection{Instruction and Metadata Format}\label{sec:inst_meta}

The accelerator uses a compact instruction format so that the same instruction stream can be broadcast to all HSF engines. As shown in \refFigure{fig:bitfield}, instructions are divided into two classes according to the most significant bit, \texttt{Class[31]}: operation instructions (O-type) and control instructions (C-type).

When \texttt{Class[31]=0}, the instruction is an O-type entry. It contains a cross-gate enable bit \texttt{CGE[30]}, a gate opcode \texttt{Opcode[29:26]} defined over $\mathcal{B}_\text{exec}$, the target-qubit index \texttt{TQI[25:20]}, the control-qubit index \texttt{CQI[19:} \texttt{14]}, and a cut-rank-table entry \texttt{CRTE[13:0]}. The \texttt{CGE} bit allows the controller to distinguish local gates from cross-boundary gates. For local instructions, the opcode and qubit fields are sufficient to drive the two Schr\"odinger Cores. For cross-boundary instructions, \texttt{CRTE} indexes the cut-rank table, which provides the rank information required by the HSF path scheduler. 

When \texttt{Class[31]=1}, the instruction is a C-type entry. It contains a control opcode \texttt{COP[30:29]} and a phase-immediate field \texttt{PI[28:0]}. C-type instructions are used for stream-control events and for delivering the immediate parameters required by phase and rotation operations. 
In particular, a C-type instruction supplies the immediate operand consumed by the next parameterized operation instruction in the stream.
This separation keeps the format compact while covering all gates in  $\mathcal{B}_\text{exec}$.

The Metadata File stores two compiler-generated structures. The first is the cut-rank table, which records the Schmidt rank of every executable cross-boundary gate. The second is the index map, which stores the final logical-to-physical permutation emitted by the HSF-S compiler. 
The index map is consulted only when logical basis queries are translated to the physical ordering expected by the compiled circuit.
Together, the instruction stream, the cut-rank table, and the index map provide a clean interface between compiler and hardware: the compiler is free to transform the circuit and qubit ordering, while the accelerator consumes only normalized instructions and associated metadata.


\vspace*{-0.2em}
\section{Experimental Evaluation}
\label{sec:exp}
\subsection{Experimental Setup}
\label{sec:ex_set}

We evaluate HSF-S from both software-simulation and hardware perspectives. On the software side, we use Google's qsim library~\cite{QuantumAI:qsim} through the qsimcirq interface. HSF path execution is performed with qsimh, while full state-vector qsim is used for reference amplitude validation. The software stack consists of qsimcirq v0.22.0 and Cirq v1.6.1.

The benchmark suite consists of three groups. From QASMBench~{\cite{QASMBench}}, we select three circuits: \texttt{qft\_n18}, \texttt{qft\_n29}, and \texttt{bigadd}- \texttt{er\_n18}. From MQT Bench~{\cite{MQTBench}}, we generate \texttt{qft}, \texttt{qpeexact}, \texttt{draper} \texttt{\_qftAdder}, \texttt{vqe\_two\_local} (\texttt{entanglement=full}, \texttt{reps=1}), and \texttt{qaoa} (\texttt{repetitions=1}, \texttt{seed=10}) for
$
n \in \{18,20,22,24,26,28,30,32\},
$
and add \texttt{randomcircuit\_n18}, yielding 41 circuits. The third group is the 12-circuit QAOA-SBM benchmark based on the weighted MaxCut construction of Herzog et al.~\cite{Herzog:DAC25}.
In this set, qubit counts are 30, 31, 32, and 33; graphs are generated with \texttt{networkx.stochastic\_} \texttt{block\_model}; all circuits are depth-$p{=}1$ weighted MaxCut QAOA; and edge weights are sampled from $\mathrm{Uniform}(0,2\pi)$ with fixed seeds. 
We fix $p_{\mathrm{intra}}=0.80$, and use
$
p_{\mathrm{inter}} \in \{0.10,0.15,0.17\}
$
for the 30- and 31-qubit cases and
$
p_{\mathrm{inter}} \in \{0.10,0.11,0.12\}
$
for the 32- and 33-qubit cases. 
In total, the benchmark suite contains 56 circuits. For all circuits, the bipartition boundary is set to
$
b=\lfloor n/2 \rfloor
$.

These benchmark families were chosen because they induce long-range or dense two-qubit interactions that are unfavorable to HSF. QFT, Draper QFT adder, and qpeexact contain long-range controlled-phase structures, while the VQE benchmark uses a fully entangled TwoLocal ansatz and the QAOA/QAOA-SBM benchmarks require entangling interactions for graph edges, which can rapidly increase cross-partition interactions in dense instances~\cite{Cleve:ProcRSocA98, Draper:arXiv00, Farhi:arXiv14, Sim:AdvQT19, Jin:SC24, Herzog:DAC25}.

As described in Section~\ref{sec:pipeline}, HSF-S sweeps parameters in Stage~2 and 3. Both stages use fraction-based parameterization with respect to the total number of two-qubit gates $|G^{2q}|$, using the common set
$
\mathcal{F} = \{0.02, 0.05, 0.1, 0.25, 0.5, 0.75, 1.0\}.
$
For each $f \in \mathcal{F}$, the corresponding window length is
$
L_f = \max\bigl(2,\mathrm{round}(f\cdot |G^{2q}|)\bigr),
$
with duplicate integer values removed after rounding. Stage~2 uses these candidates for $L_{\mathrm{reorder}}$, and Stage~3 uses the same set for $L_{\mathrm{swap}}$. For discounted SWAP insertion, the half-life parameter is chosen from
$
H \in \{0.25L,\;0.5L,\;1.0L\},~~ H \leftarrow \max(0.5,H),
$
again with duplicates removed. Each $H$ defines a discount factor $\gamma$, and the undiscounted baseline $\gamma=1$ is additionally included. For the Stage~2 probe and the ablation study below, the default SWAP setting is
$
\bigl(0.5|G^{2q}|,\;\gamma=1\bigr),
$
\PSC{and Stage~3 uses a candidate cap of 
$
P=\mathrm{min}(b,12).
$}

\begin{figure}[t]
\centering
\includegraphics[width=\columnwidth]{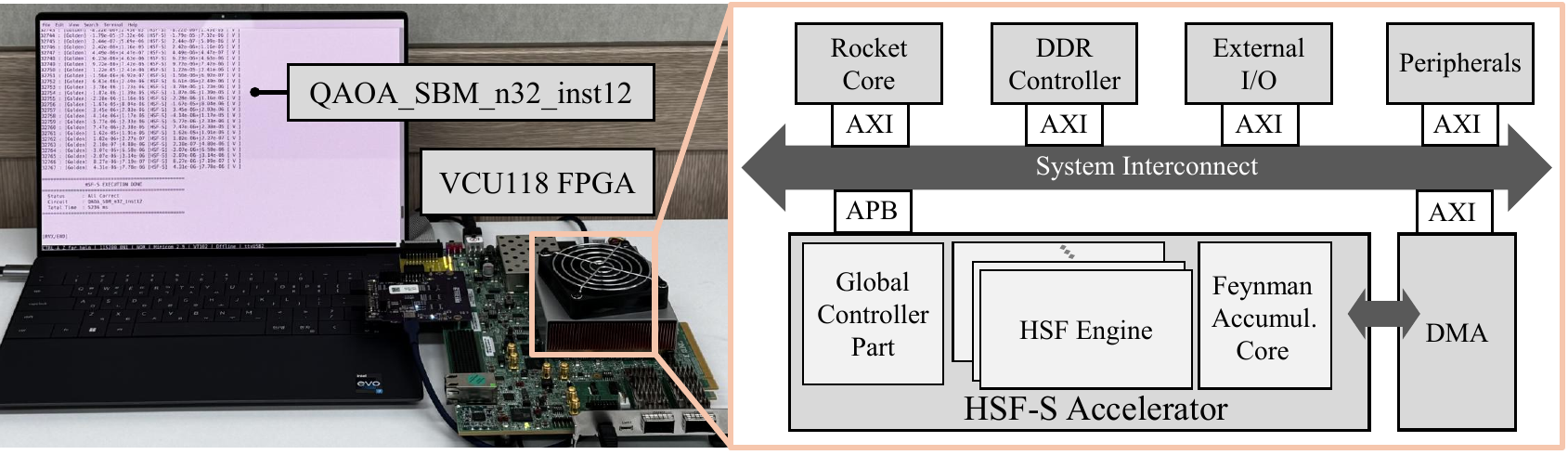}
\vskip -6pt
\caption{Prototype demonstration of the stand-alone HSF-S QC emulator on the AMD VCU118 FPGA evaluation board.}
\label{fig:VCU118}
\end{figure}

\begin{table}[t]
\caption{FPGA resource utilization of the HSF-S processor prototype and post-synthesis area/power results (UMC 14\,nm, 200\,MHz).}
\vskip -9pt
\centering
\resizebox{1\columnwidth}{!}{
          \renewcommand{\arraystretch}{1.2}
          \begin{tabular}{ l| r r r | r r}
            \Xhline{1pt}
            \rowcolor[HTML]{D9D9D9}
              IPs & LUTs & FFs & DSPs & Area [$\mu$m$^2$] & Power [mW] \\ \hline
              Rocket Core             & 15500  & 9775  & 4   & 62{,}917  & 4.29 \\
              DDR Controller             & 17131  & 21246  & 3   & 40{,}942  & 3.35 \\
              Peripherals incl. I/O       & 15442  & 17417  & 0   & 32{,}975  & 2.09 \\
              System Interconnect     & 9598  & 15111  & 0   & 26{,}520  & 1.50 \\
            \Xhline{0.6pt}
\rowcolor[HTML]{EFEFEF}
           \textbf{HSF-S Accelerator}  & \textbf{911{,}195} & \textbf{720{,}825} & \textbf{5{,}312} &
  \textbf{21{,}753{,}709} & \textbf{601.96} \\
             $\llcorner$ Global Control Part$^{\dagger}$                   & 46{,}691 & 16{,}153 & 0
   & 1{,}288{,}040 & 36.15 \\
             $\llcorner$ HSF Engine ($\times16$)                           & 52{,}828 & 42{,}973 & 332
   & 1{,}277{,}316 & 34.96 \\
            \quad $\llcorner$ Local Controller                                 & 2{,}433  & 3{,}113  & 0
   & 1{,}529 & 0.53 \\
            \quad $\llcorner$ Schr\"odinger Core$^{\ddagger}$ ($\times2$)                      & 21{,}751 & 13{,}753 & 134
   & 624{,}141 & 16.07 \\
            \quad $\llcorner$ Feynman Comp. Core                             & 6{,}893  & 12{,}354 & 64
   & 27{,}505 & 2.29 \\
             $\llcorner$ Feynman Acc. Core                           & 19{,}256 & 17{,}104 & 0
   & 28{,}613 & 6.45 \\
          \Xhline{1pt}
          \multicolumn{6}{r}{\normalsize $^{\dagger}$\,includes 8 RAMB36 and 32 URAM288;
  $^{\ddagger}$\,includes 16 URAM288 for state memory.}
          \end{tabular}
          }
\label{tab:fpga_and_ASIC_breakdown}
\end{table}

\LKW{For runtime measurements, we query amplitudes for logical indices from $0$ to $2^{15}-1$ for every compiled circuit and fix the qsimh thread count to 8. Since HSF is query-driven, requesting all $2^n$ amplitudes is unnecessary; using $2^{15}$ queries provides a common workload that is large enough to expose runtime trends while keeping execution practical across the benchmark suite. All qsimh executions use a uniform wall-clock timeout of 1 hour. Runs that do not finish within this budget are marked as \texttt{Timeout} (1h).
Since prior HSF studies primarily report CPU-based results~\cite{Burgholzer:QCE21,Im:ASPDAC23,Herzog:DAC25}, we use a CPU platform as the main software baseline. These software-side preprocessing and simulator experiments are performed on an Intel Core Ultra 9 285K (24 physical cores, efficient cores disabled, 3.7\,GHz, 125\,W) running Ubuntu 24.04 LTS.}

On the hardware side, we implement the proposed HSF-S accelerator in Verilog RTL and integrate it into a stand-alone HSF-S processor centered on the Rocket RISC-V core~\cite{Rocket}. The processor also includes external DDR4 memory, a lightweight NoC~\cite{Han:TCAD18}, and I/O peripherals. 
An EDA tool, RISC-V eXpress (RVX) widely adopted for developing processors on the RISC-V platform~\cite{Han:IoTJ2021,Park:TCASI24,Lee:IoTJ25,Jeon:DAC25,Cho:DATE2026,Kwak:DATE26}, is used for the SoC integration.
We prototype the integrated design on the AMD VCU118 FPGA evaluation board~\cite{VCU118} at 200\,MHz. To fit the FPGA resource budget, the instantiated accelerator contains 16 HSF engines and 8 processing units per Schr\"odinger core\LKW{, all operating on 64-bit complex arithmetic}. As shown in Figure~\ref{fig:VCU118}, the host PC runs the preprocessing flow of Section~\ref{sec:pipeline}, generates the compiled instruction stream, index map, and cut-rank table, and transfers them to the HSF-S processor, which then executes the HSF workload. After FPGA validation, we synthesize the design using Synopsys Design Compiler~\cite{DesignCompiler} with a UMC 14\,nm \LKW{technology} to obtain area and power estimates. \refTable{tab:fpga_and_ASIC_breakdown} reports the FPGA resource breakdown together with the post-synthesis results.
The 32 Schrödinger cores (16 engines $\times$ 2 cores), each containing dedicated state memory and 8 processing units, account for the largest share of both FPGA and post-synthesis resources.

\subsection{Experimental Results}
\label{sec:ex_result}

\begin{figure*}[t]
\centering
\hspace{-12pt}
\includegraphics[width=1.02\linewidth]{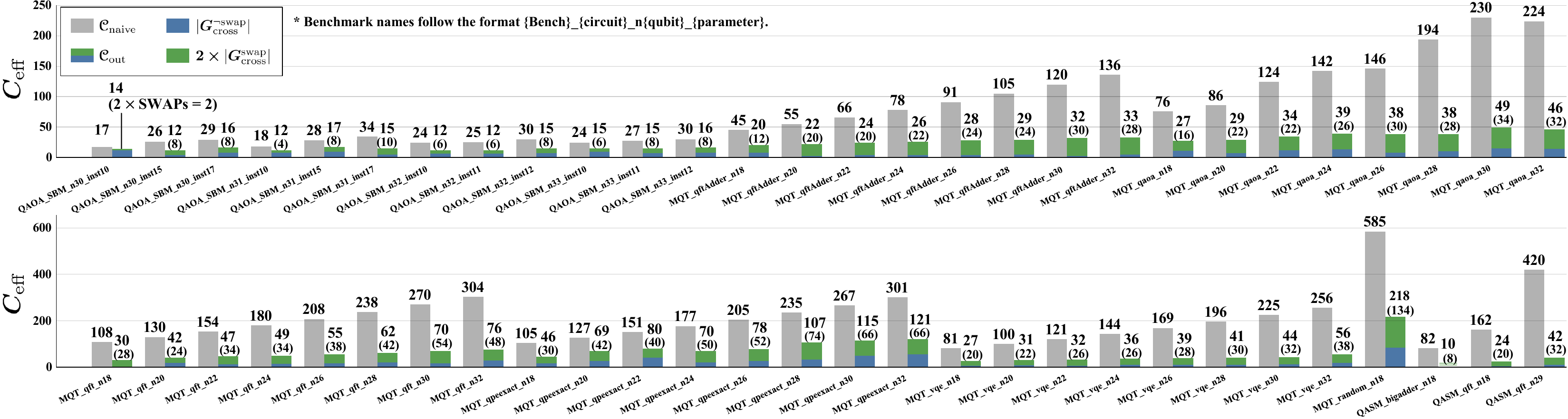}
\vskip -10pt
\caption{Reduction in effective HSF path cost before and after HSF-S preprocessing. Gray bars denote the naive $\mathbf{C_\text{eff}}$, blue bars denote the remaining non-SWAP cross-boundary contribution $\mathbf{|G_\text{cross}^{\neg\text{swap}}|}$ after optimization, and green bars denote the cross-SWAP contribution $\mathbf{2|G_\text{cross}^{\text{swap}}|}$.}
\label{fig:cost_results}
\end{figure*}

\begin{figure}[t]
\vskip -3pt
\centering
\includegraphics[width=1\columnwidth]{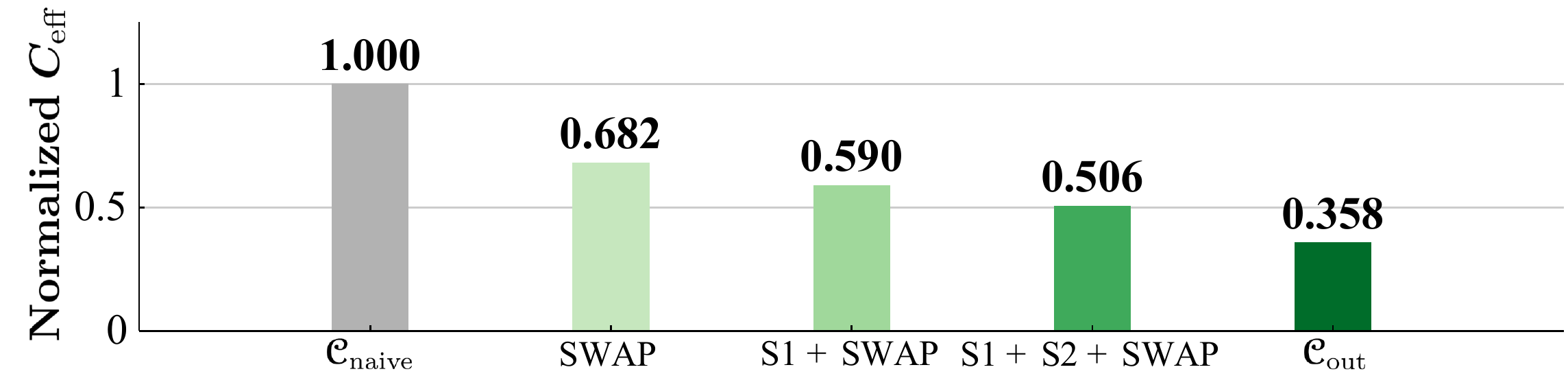}
\vskip -6pt
\caption{Ablation of the HSF-S pipeline. Normalized aggregate $\mathbf{C_\text{eff}}$ decreases monotonically as additional stages are enabled.}
\label{fig:ablation}
\end{figure}

We evaluate HSF-S using two metrics: the effective path cost $C_\text{eff}$ and execution runtime. Before discussing performance, we first confirm functional correctness. For all evaluated circuits and all queried logical basis indices, the HSF-S-compiled circuits agree with the corresponding naive-lowered reference amplitudes up to floating-point numerical precision after index-map translation, with a worst-case maximum absolute error of $3.37\times10^{-8}$.

\refFigure{fig:cost_results} compares $C_\text{eff}$ before and after HSF-S preprocessing. Averaged over benchmark groups, HSF-S reduces $C_\text{eff}$ by 87.66\% on the QASMBench subset, 68.48\% on MQT Bench, and 43.79\% on QAOA-SBM. The largest per-circuit reduction is observed on \texttt{QASM\_qft\_n29}, where $C_\text{eff}$ decreases from 420 to 42 (90.00\%), while the smallest is observed on \texttt{QAOA\_SBM\_n30\_inst10}, where $C_\text{eff}$ decreases from 17 to 14 (17.65\%). 
Overall, circuits with larger naive $C_\text{eff}$ tend to admit larger reductions, because selective SWAP insertion can eliminate more repeated cross-boundary gates.

On the QAOA-SBM benchmark set, HSF-S achieves path-reduction factors ranging from $2^3\times$ to $2^{19}\times$, with a median of $2^{13}\times$. This is close to the min/median/max path reductions of $2^6\times$, $2^{13}\times$, and $2^{20}\times$ reported for joint cutting~\cite{Herzog:DAC25}. Unlike joint cutting, however, HSF-S does not rely on the presence of only a small number of contiguous cross-boundary blocks, and therefore remains applicable to a broader range of circuit structures.


\refFigure{fig:ablation} isolates the contribution of each pipeline stage. We compare five configurations: the naive baseline, a SWAP-only baseline, Stage~1 with default SWAP, Stage~1+2 with default SWAP, and the full HSF-S pipeline. For each circuit, $C_\text{eff}$ is first normalized by its naive value, then aggregated across the benchmark suite, and finally renormalized by the aggregate naive cost. The resulting normalized costs are 1.000, 0.682, 0.590, 0.506, and 0.358, respectively. The monotonic decrease shows that the gains from SWAP tuning do not arise in isolation; rather, Stage~1 and Stage~2 first reshape the gate order so that later SWAP insertion becomes more effective.

\begin{figure}[t]
\centering
\includegraphics[width=\columnwidth]{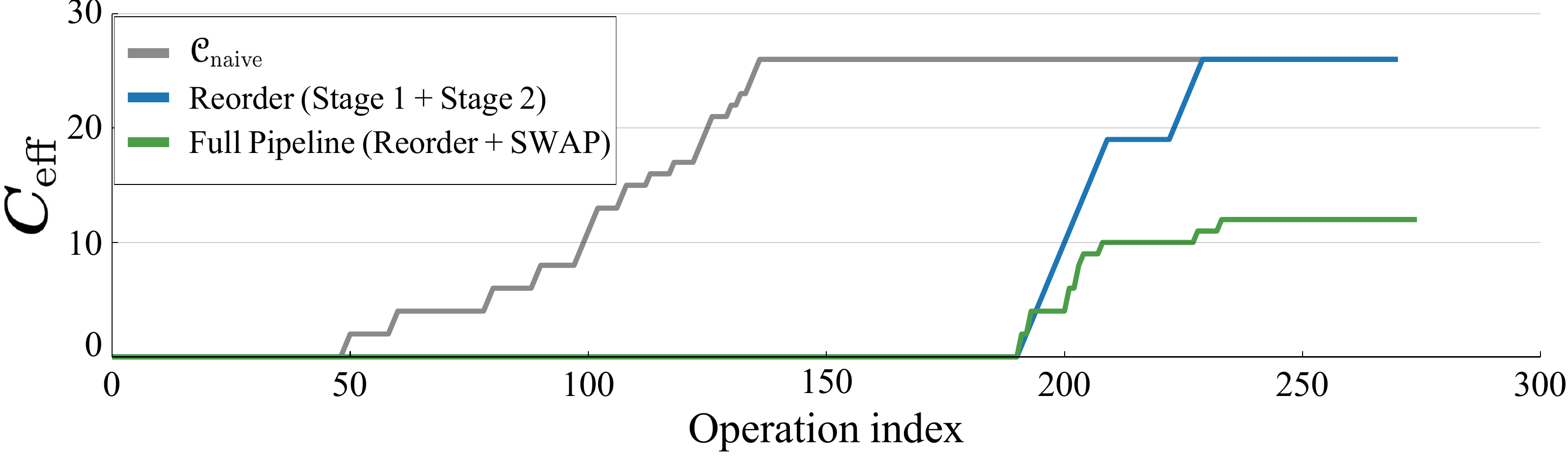}
\vskip -8pt
\caption{Cumulative effective path cost along the circuit order for a representative QAOA-SBM circuit across HSF-S pipeline stages.}
\label{fig:cummulative}
\end{figure}

\refFigure{fig:cummulative} provides a time-axis view of this effect on a representative circuit, \texttt{QAOA\_SBM\_n30\_inst15}. After reordering, the dominant cross-boundary interactions become more concentrated in later regions of the circuit rather than being dispersed throughout the full schedule. Subsequent SWAP insertion directly suppresses these clustered cost spikes and lowers the cumulative cost. Thus, HSF-S reduces not only the total $C_\text{eff}$, but also restructures where the dominant branching burden appears.

To evaluate whether the reduction in $C_\text{eff}$ translates into practical speedup, \refTable{tab:runtime} compares qsimh runtime before and after HSF-S preprocessing and reports the runtime of the HSF-S processor on the same post-preprocessing workloads. All runtimes exclude preprocessing time. Across the 56-circuit suite, full HSF-S compilation takes 2.81\,s on average, with the longest case (\texttt{MQT\_random\_n18}) taking 29.2\,s. 
\LWJ{For completeness, we also evaluated both naive and HSF-S-preprocessed workloads on an NVIDIA GeForce RTX 5090 GPU (32\,GB VRAM) under the same 1-hour timeout budget, but none of the benchmark circuits completed within this limit; GPU runtimes are therefore omitted from \refTable{tab:runtime}. The results reveal two complementary effects.}


First, HSF-S preprocessing consistently reduces software runtime whenever it lowers $C_\text{eff}$. For example, \texttt{QAOA\_SBM\_n30\_inst15} drops from timeout to 0.652\,s after preprocessing, and \texttt{QASM\_bigadder\_n18} drops from timeout to 0.077\,s. These measurements are also consistent with the HSF runtime model in Section~\ref{sec:hsf}: for circuits of the same size and under the same query budget, runtime scales approximately with $2^{\Delta C_\text{eff}}$, reflecting the dominant effect of path branching. For circuits with similar $C_\text{eff}$, larger qubit counts increase the per-path workload $W_\text{path}$ and therefore also increase runtime. This confirms that reducing recurring cross-boundary interactions directly improves the practical tractability of HSF execution and supports the use of $C_\text{eff}$ as an effective compile-time surrogate for practical HSF runtime.

Second, the hardware implementation provides an additional speedup on top of the preprocessing gain. Among the eight circuits completed by both CPU-preprocessed software and the HSF-S hardware, the proposed accelerator achieves 1.87$\times$ to 4.34$\times$ speedup, with an average of 2.94$\times$. For instance, \texttt{QAOA\_SBM\_n30\_inst10} is reduced from 4.550\,s to 1.049\,s, \texttt{QAOA\_SBM\_n32\_inst12} from 20.196\,s to 5.236\,s, and \texttt{QASM\_bigadder\_n18} from 0.077\,s to 0.020\,s. 
Moreover, for \texttt{MQT\_qaoa\_n18} and \texttt{MQT\_vqe\_n18}, CPU-preprocessed qsimh still times out at 1 hour, whereas the HSF-S hardware completes the same post-preprocessing workloads in 2484.348\,s and 2425.152\,s, corresponding to at least 1.45$\times$ and 1.48$\times$ speedup, respectively. These results show that HSF-S preprocessing and hardware acceleration are complementary: preprocessing reduces the exponential branching burden, while the accelerator further lowers the execution overhead of the resulting path workload through an HSF-specific execution engine.

\begin{table}[t]
\centering
\scriptsize
\caption{Effective HSF cost and runtime before and after HSF-S preprocessing on representative circuits (excluding preprocessing time).}
\vskip -8pt
\label{tab:runtime}
\setlength{\tabcolsep}{5pt}
\renewcommand{\arraystretch}{1.2}
\begin{tabular}{lccccc}
\toprule
& \multicolumn{2}{c}{Cost ($C_{\mathrm{eff}}$)} & \multicolumn{3}{c}{Runtime (s)} \\
\cmidrule(lr){2-3} \cmidrule(lr){4-6}
Circuit & $\mathcal{C}_{\mathrm{naive}}$ & $\mathcal{C}_{\mathrm{out}}$ & Naive & HSF-S & HSF-S (HW) \\
\midrule
QASM\_qft\_n18         & 162 & 24 & Timeout (1h) & 693.380      & 371.270  \\
QASM\_bigadder\_n18    &  82 & 10 & Timeout (1h) & 0.077        & 0.020    \\
MQT\_qaoa\_n18         &  76 & 27 & Timeout (1h) & Timeout (1h) & 2484.348 \\
MQT\_vqe\_n18          &  81 & 27 & Timeout (1h) & Timeout (1h) & 2425.152 \\
QAOA\_SBM\_n30\_inst10 &  17 & 14 & 36.200       & 4.550        & 1.049    \\
QAOA\_SBM\_n30\_inst15 &  26 & 12 & Timeout (1h) & 0.652        & 0.348    \\
QAOA\_SBM\_n30\_inst17 &  29 & 16 & Timeout (1h) & 14.507       & 6.500    \\
QAOA\_SBM\_n32\_inst10 &  24 & 12 & Timeout (1h) & 1.267        & 0.603    \\
QAOA\_SBM\_n32\_inst11 &  25 & 12 & Timeout (1h) & 2.077        & 0.616    \\
QAOA\_SBM\_n32\_inst12 &  30 & 15 & Timeout (1h) & 20.196       & 5.236    \\
\bottomrule
\end{tabular}
\end{table}


\section{Conclusion}
This paper presented HSF-S, a compiler--accelerator co-designed framework for exact HSF-based quantum circuit emulation. HSF-S combines HSF-compatible gate lowering, a rank-aware effective path-cost model, dependency-preserving reordering, and discounted-gain SWAP insertion to reduce recurring cross-boundary interactions while preserving exact semantics and query correctness. We further designed a dedicated HSF-S accelerator and metadata-driven execution flow, and validated the resulting stand-alone RISC-V-based system through FPGA prototyping and 14\,nm synthesis. 
Across the evaluated benchmarks, HSF-S matched the reference amplitudes up to floating-point precision, reduced $C_\text{eff}$ by up to 90.0\%, and substantially improved practical tractability. Furthermore, the hardware accelerator provided an additional speedup of up to 4.34$\times$ over HSF-S-preprocessed software running on a CPU. These results show that compiler-guided path reduction and HSF-specific hardware support make exact HSF emulation substantially more scalable and practical on resource-constrained platforms.

\bibliographystyle{unsrt}
\bibliography{ref}

\end{document}